\documentclass[%
 aip,
 amsmath,amssymb,
 preprint,%
]{revtex4-1}\usepackage{setspace}
\usepackage{graphicx}
\usepackage{dcolumn}
\usepackage{bm}

\usepackage[utf8]{inputenc}
\usepackage[T1]{fontenc}
\usepackage{mathptmx}
\newcommand{\newaffinity}{density-threshold affinity}
\newcommand{\newaffinities}{density-threshold affinities}
\newcommand{\Newaffinities}{Density-threshold affinities}

\newcommand{\nachr}{nAChR}
\newcommand{\plgic}{pLGIC}
\newcommand{\xo}{\textit{Xenopus} oocytes}
\usepackage{graphicx}
\usepackage{mathtools}
\usepackage{color}
\usepackage{lineno}

\usepackage{amssymb}

\begin{document}




\title{Spontaneous Lipid Binding to the Nicotinic Acetylcholine Receptor in a Native Membrane}


%
\preprint{AIP/123-QED}

\author{Liam Sharp}
 \affiliation{Center for Computational and Integrative Biology, Rutgers University-Camden, Camden, NJ}
\author{Grace Brannigan}
\affiliation{Center for Computational and Integrative Biology, Rutgers University-Camden, Camden, NJ}
 \affiliation{Department of Physics, Rutgers University-Camden, Camden, NJ}

\begin{abstract}
The nicotinic acetylcholine receptor (nAChR) and other pentameric ligand-gated ion channels (pLGICs) are native to neuronal membranes with an unusual lipid composition. While it is well-established that these receptors can be significantly modulated by lipids, the underlying mechanisms have been primarily studied in model membranes with only a few lipid species. Here we use coarse-grained molecular dynamics (MD) simulation to probe specific binding of lipids in a complex quasi-neuronal membrane. We ran a total of 50 microseconds of simulations of a single nAChR in a membrane composed of 36 species of lipids. Competition between multiple lipid species produces a complex distribution. We find that overall, cholesterol selects for concave intersubunit sites and PUFAs select for convex M4 sites, while monounsaturated and saturated lipids are unenriched in the nAChR boundary. In order to characterize binding to specific sites, we present a novel approach for calculating a ``density-threshold affinity'' from continuous density distributions. We find that affinity for M4 weakens with chain rigidity, which suggests flexible chains may help relax packing defects caused by the conical protein shape. For any site, PE headgroups have the strongest affinity of all phospholipid headgroups, but anionic lipids still yield moderately high affinities for the M4 sites as expected.  We observe cooperative effects between anionic headgroups and saturated chains at the M4 site in the inner leaflet. We also analyze affinities for individual anionic headgroups. Combined, these insights may reconcile several apparently contradictory experiments on the role of anionic phospholipids in modulating nAChR.
\end{abstract}

%
\maketitle



\section{Introduction}
\label{Intro}

The nicotinic acetylcholine receptor (\nachr)~is a well studied excitatory pentameric ligand gated ion channel (\plgic s). \nachr s are found at high density in post-synaptic membranes and the neuromuscular junction in mammals, and the electric organ in \textit{Torpedo} electric rays. The \nachr~is activated by the binding of agonists such as nicotine or acetylcholine to the orthosteric site in the extra-cellular domain (ECD). When post-synaptic \nachr s are activated {\it en-mass} they stimulate an action potential. Thus \nachr s~play a critical role in both cognition and memory\cite{Henault2015} and neuromuscular function \cite{Mukhtasimova2016,Kalamida2007}. \nachr~ and the greater \plgic~superfamily play various roles in neurological diseases related to inflammation \cite{Taly2009,Patel2017,Yocum2017,Egea2015},  addiction \cite{Cornelison2016}, chronic pain \cite{Xiong2012}, Alzheimer's Disease \cite{Walstab2010,Picciotto_Neuroprotection_2008,MartinRuiz_4_1999,Kalamida2007},spinal muscular atrophy \cite{Arnold_Reduced_2004}, schizophrenia \cite{Haydar2010,Kalamida2007} and neurological autoimmune diseases \cite{Lennon_Immunization_2003, Kumari2008}.

\nachr s~are highly sensitive to their local lipid environment. \nachr~ poorly conducts ions in model phosphatidylcholine (PC)-only membranes, but can conduct a current with the addition of cholesterol or anionic lipids \cite{Baenziger2017,Dalziel1980,Ellena1983,Criado1983,Fong1986,Fong1987,Jones1988,Sunshine1994,DaCosta2009b,Baenziger2017}, though too much  cholesterol can also cause a loss of function \cite{Criado1983,Mantipragada2003,Barrantes2010a,Baier2011a}. 
Functional studies using \xo~ \cite{Zhou2003,Gamba2005,Chen2015,Kouvatsos2016,Nys2016,Polovinkin2018,Moffett2019,Kumar2020} require lipid additives such as asolectin\cite{Criado1983,Zhou2003,Gamba2005,Chen2015,Nys2016,Polovinkin2018,Moffett2019,Kumar2020} or lipids from synaptic membranes \cite{Conti2013} to recover native levels of \nachr{} ion flux.  
Understanding the mechanism of modulation requires understanding how and where the modulating lipid interacts with the receptor, and these interactions may themselves be dependent upon the rest of the lipid composition.  

Mammalian neuronal membranes \cite{Isolated1969, Taguchi2010, Breckenridge1973,Ingolfsson2017b} have unique compositions compared to other mammalian membranes\cite{McEvoy2000,Kim2001,VanMeer2010,Lorent2020,Ingolfsson2014}. Neuronal membranes are more similar to the membrane of the \textit{Torpedo} electric ray's electric organ \cite{Barrantes1989,Quesada2016} than the average mammalian membrane\cite{Ingolfsson2014}. The neuronal membrane \cite{Isolated1969, Taguchi2010, Breckenridge1973,Ingolfsson2017b} is rich in lipids in which one or both chains are polyunsaturated fatty acids (PUFAs), particularly the $n-6$ PUFA arachidonic acid (AA), and the $n-3$ PUFAs docosahexaenoic acid (DHA) and eicosapentaenoic acid (EPA). These three PUFA's comprise $\sim20-25\%$ of the acyl chains of neuronal phospholipids, and are involved in secondary signaling \cite{McNamara2006,McNamara2008} and neuronal development \cite{Maekawa2017}. PUFAs are linked to a number of neurological diseases and disorders that overlap \nachr~related diseases. PUFAs play a roll in major depressive and bipolar disorder \cite{MuralikrishnaAdibhatla,McNamara2008,Schneider2017,Koga2019,Hamazaki2015}, schizophrenia \cite{Peet2003,Bushe2005,Berger2006,Schneider2017,Maekawa2017,Hamazaki2015}, and Alzheimer's Disease \cite{Conquer2000,DiPaolo2011,Bennett2013,MuralikrishnaAdibhatla,Yadav2014a,Escriba2017}. 

Functional experiments have focused on the role of anionic lipids and cholesterol as modulators of \plgic s  \cite{Ellena1983,Fong1986,Fong1987,Jones1988,Sunshine1994,DaCosta2009b,Dalziel1980,Addona1998,Criado1983} (the role of polyunsaturation has received comparatively little attention due to common challenges with oxidation of polyunsaturated chains). Such experiments have been overwhelmingly consistent with a role for direct binding of lipids as a modulatory mechanism.   
As for most membrane proteins, it is experimentally challenging to capture the boundary lipid composition of \plgic~because lipids are small molecules that may remain partly fluid even in their bound state. Numerous structures of \plgic s have revealed a conserved arrangement for both the TMD and the ECD. In the TMD, each subunit has four membrane helices (M1-M4) with the five subunits forming a ``star'' shape around a central pore (Figure \ref{fig:PBT}A). The M2 helix lines the pore, the M1 and M3 helices form a middle ring that includes the intersubunit cavities, and the M4 helices form the tip of the star.  Structural methods have resolved potential cholesterol molecules\cite{Laverty2017, Budelier2019} and phospholipids \cite{Basak2017, Henault2019, Kim2020} bound to subunit interfaces, but crystallographic disorder introduced by lipids typically precludes identification of lipid species. Mass spectrometry has revealed specific binding of anionic lipids, with additional mutagenesis studies suggesting localized sites in the inner leaflet near the M4 helices. \cite{Tong2019} 

Molecular dynamics (MD) simulations are particularly useful for visualizing and characterizing microscopic interactions within a fluid system. Given a putative cholesterol or lipid binding mode, atomistic simulations can be used to probe stability of the lipid binding mode.  For pentameric channels, such approaches have primarily demonstrated stability of bound cholesterol\cite{Brannigan2008}, particularly at intersubunit sites\cite{Laverty2017,Henin2014a}.  Unfortunately, fully atomistic simulations suffer from slow diffusion of lipids within the membrane, which prevents spontaneous lipid sorting by proteins over accessible simulation time scales. 

Coarse-grained MD simulations use simplified molecular models that can reveal spontaneous lipid sorting, domain formation, and protein partitioning over simulation timescales \cite{Domanski2012,Chavent2016,Carpenter2018,Ingolfsson2020}. Coarse-grained MD simulations have been used previously to probe interactions of \plgic s with propofol\cite{Joseph2016} as well as spontaneous lipid binding in model membranes\cite{Sharp2019,Woods2019,Tong2019}.  
In previous work, we found that \nachr~embedded in multiple domain-forming model membranes partitioned to the PUFA-rich liquid disordered domain\cite{Sharp2019}, rather than to the cholesterol-rich liquid-ordered or ``raft'' domain that was suggested by cholesterol modulation.  We observed that cholesterol still occupies embedded sites on the \nachr{} TMD, where it is shielded from the liquid disordered domain. However, native membranes are primarily composed of heteroacidic lipids with two distinct chains, where each chain has a different domain preference; such lipids will naturally destabilize domains. In non-domain forming model membranes composed of heteroacidic lipids, two classes of five-fold symmetric sites emerged: an intersubunit site and the M4 site (Figure \ref{fig:PBT}B).  Cholesterol and saturated chains were enriched at the~inter-subunit interfaces and n-3 PUFA acyl-chains were enriched around the M4 helices\cite{Woods2019}. These results were consistent with binding to minimize packing defects: the rigid lipids could fill in the concave regions at the intersubunit sites while the flexible chains would easily deform around the ``star points'' of the M4 helices.  Yet it was not clear whether these same patterns would be upheld in the more complex environment of a native neuronal membrane, which has many more options for minimizing any packing defect.  

Neuronal membranes also contain a sizeable fraction of anionic lipids in the inner leaflet\cite{Isolated1969,Breckenridge1973,Taguchi2010}. With collaborators in the Cheng lab, we recently\cite{Tong2019} showed that anionic headgroups bind preferentially to the \plgic{} Erwinia ligand-gated ion channel (ELIC), when the same acyl chains are used for both headgroups.  Through coarse-grained MD, we found specific binding sites for 1-palmitoyl-2-oleoyl phosphatidylglycerol POPG in the intersubunit sites (inner leaflet); these sites contained basic amino acids that were also implicated through mutagenesis\cite{Tong2019}.  In \nachr{} the high-density of basic amino acids are in the M4 site (inner leaflet) rather than the intersubunit site (inner leaflet), so we would expect a shift for \nachr{} even in model membranes, due purely to the protein sequence. 
The relative roles of headgroup charge {\it vs} acyl chain saturation in driving affinity are unknown. 

The use of complex quasi-realistic membranes in coarse-grained MD simulations is growing more feasible. In 2014, Ing{\'o}lfsson et al\cite{Ingolfsson2014} simulated an ``average mammalian'' membrane containing 63 lipids species, followed in 2017 by a coarse-grained neuronal membrane \cite{Ingolfsson2017b}. Multiple accessible and realistic membranes have been developed for comparison of protein-lipid interactions between model and quasi-native membranes\cite{Marrink2019,Wilson2020,Ingolfsson2020,Carpenter2018,Lorent2019}. To our knowledge, no such coarse-grained MD simulations using quasi-native membranes have been used with \plgic s.

While the model membranes we used previously are useful for identifying putative sites, they have critical limitations.  As stated previously, model membranes typically vary headgroup charge or acyl chain saturation, not both. Model membranes also do not allow for identification of more specific chemical variations within general saturation classes (i.e. n-3 PUFAs like DHA {\it vs} n-6 PUFAs like $\alpha$-linolenic acid) or like-charged head groups (PC {\it vs} PE, or phosphoserine (PS) {\it vs} phosphoinositol (PI)). For this work, we embed the neuromuscular \nachr\cite{Unwin2005} in a coarse-grained neuronal membrane \cite{Ingolfsson2017b}. To test whether the predictions we developed from model membranes hold for native membranes, we develop a new method for quantifying affinities for partially-occupied binding sites.

 The remainder of this paper is organized as follows. Section II presents our simulation and analysis approach, including introduction of the \newaffinity.  Section III presents results and discussion of site selectivity of neutral lipids, followed by a reoriented discussion of the same data that is focused on lipid preferences of individual sites. We then consider selectivity of anionic lipids in the inner leaflet and finally consider the effects of specific headgroup differences.  Section IV concludes.  


\section{Methods}
\label{lab}

\subsection{Simulation Composition}
All simulations used the coarse-grained MARTINI 2.2\cite{DeJong2012} topology and forcefield.
\nachr~ coordinates were based on a cryo-EM structure of the $\alpha{\beta}\gamma\delta$ muscle-type receptor in native torpedo membrane (PDB 2BG9\cite{Unwin2005}). This is a medium resolution structure (4\AA) and was further coarse-grained using the martinize.py script; medium resolution is sufficient for use in coarse-grained simulation, and the native lipid environment of the proteins used to construct 2BG9 is critical for the present study. The secondary, tertiary and quaternary structure in 2BG9 was preserved via soft backbone restraints during simulation as described below, so any inaccuracies in local residue-residue interactions would not cause instability in the global conformation.  

\nachr~was embedded in a coarse-grained neuronal membrane based on Ing{\'o}lfsson et al\cite{Ingolfsson2017b}. The neuronal membrane from described by Ing{\'o}lfsson contains phospholipids, sterols, diacylglycerol, and ceramide. Membranes presented in this paper only consider phospholipids and cholesterol, for a total of 36 unique lipid species, see Table SI 1.

Coarse-grained membranes were built using the MARTINI script insane.py \cite{Wassenaar2015}, which was also used to embed the coarse-grained \nachr~within the membrane. The insane.py script randomly places lipids throughout the inter- and extra-cellular leaflets, and each simulation presented in this manuscript was built separately. All simulation box sizes were $40x40x35$ nm$^3$ with  $\sim 4,500-5,000$ lipids and total $\sim450,000$ beads.

\subsection{Simulations}

Molecular dynamics simulations run using the MARTINI 2.2\cite{DeJong2012} forcefield and GROMACS\cite{Berendsen1995,Abraham2015}  2019.2 . All systems used van der Waals (vdW) and Electrostatics with reaction-field and a dielectric constant of $\epsilon_r$=15 and electrostatic cutoff length at 1.1 nm. Energy minimization was performed for 1000000 steps, but energy minimization tended to concluded after $\sim 5000-10000$ steps.

Volume and pressure equilibrations were run with isothermal-isochoric (NVT) and isothermal-isobaric (NPT) ensembles respectively. NVT and NPT simulations used a time step of 15 fs and run for 0.3 ns using Berendsen thermostat held at a temperature of 323 K, and Berendsen pressure coupling with compressibility set to 3$\times$10$^{-5}$ bar$^{-1}$ and a pressure coupling constant set to 3.0 ps  for the NPT ensemble. 

Molecular dynamics simulations were run using a time step of 20 fs for 5 $\mu$s for 10 replicas. Simulations were conducted in the NPT ensemble, by using the velocity rescaling to a temperature of 323 K with a coupling constant set to 1 ps. Semi-isotropic pressure coupling was set to Parrinello-Rahman with compressibility at 3$\times$10$^{-5}$ bar$^{-1}$ and pressure coupling constant set to 3.0 ps. 

Secondary structures restraints with MARTINI recommendations were constructed by the martinize.py \cite{DeJong2012} script and imposed by GROMACS \cite{Berendsen1995,Abraham2015}. The \nachr~conformation was preserved by harmonic bonds between backbone beads separated by less than 0.5 nm and calculated using the ElNeDyn algorithm \cite{Periole2009} associated with MARTINI \cite{DeJong2012} with a coefficient of 900 kJ$\cdot$mol$^{-1}$.
\begin{figure*}[!h]
	\center
	\includegraphics[width=\linewidth]{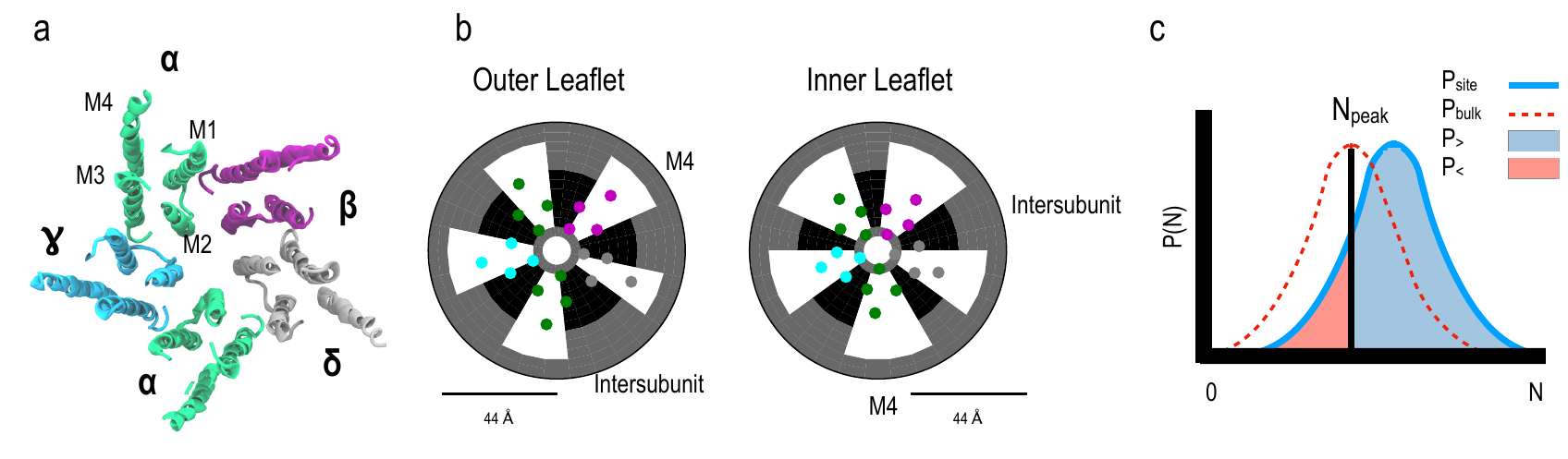}
	\caption{{ Binding site boundaries and distribution definitions.}  (a) Structure of the \nachr {}TMD\cite{Unwin2005}, viewed from the extracellular domain. Helices are colored by subunit ($\alpha$:green, $\beta$:purple, $\gamma$: cyan, $\delta$:grey). (b) Boundaries of the pseudo-symmetric intersubunit (black) and M4 (white) sites. The angular components are determined by the location of the M1 and M3 alpha-helices for either two adjacent subunits (intersubunit sites) or a single subunit (M4 sites), and are listed in Table S5. Circles correspond to the helices shown in panel A.  (c)  The distributions $P_{site} (n)$ (blue) and $P_{bulk}(n)$ (dashed red) represent the probability distributions for number of beads of a certain lipid species in the site or in an analogously-sized area of the bulk, respectively. The value $n_{peak}$ maximizes $P_{bulk}$. $P_<$ (pink) is the area under $P_{site}$ to the right of $n_{peak}$ while $P_>$ (light blue) is the area under $P_{site}$ to the right of $n_{peak}$. }
	\label{fig:PBT}
\end{figure*}

\subsection{Calculation of Polar Density Distributions}
As in our previous work\cite{Sharp2019, Woods2019, Tong2019}, the two-dimensional density distribution $\rho_{B}$ of the beads within a given lipid species $B$ around the protein was calculated on a polar grid:
  \begin{equation}
      \rho_{B}(r_i,\theta_j)= \frac{\left\langle n_{B}(r_i,\theta_j) \right\rangle}{r_i \Delta{r}\Delta{\theta}} \\        
    \label{eq:R}
  \end{equation}
  where  $r_i = i \Delta{r}$ is the projected distance of the bin center from the protein center, $\theta_j = j \Delta{\theta}$ is the polar angle associated with bin j,  $\Delta{r}$= 10\AA~ and  $\Delta{\theta} = \frac{\pi}{15}$ radians are the bin widths in the radial and angular direction respectively, and $\left\langle n_{B}(r_i,\theta_j) \right\rangle$ is the time-averaged number of beads of lipid species $B$ found within the bin centered around radius $r_{i}$ and polar angle $\theta_{j}$.  In order to determine enrichment or depletion, the normalized density $ \tilde{\rho}_{B}(r_i,\theta_j)$ is calculated by dividing by the approximate expected density of beads of lipid type B in a random mixture, $x_{B}s_{B}~N_{L}/\langle L^{2}\rangle$, where $s_{B}$ is the number of beads in one lipid of species B, $N_{L}$ is the total number of lipids in the system, and $\langle L^{2}\rangle$ is the average projected box area:
  \begin{equation}
  \tilde{\rho}_{B}(r_i,\theta_j)=\frac{ \rho_{B}(r_i,\theta_j)}{x_{B}s_{B}~N_{L}/\langle L^{2}\rangle} \\        
    \label{eq:Rt}
  \end{equation}
where the expected density is derived at the first frame of the simulation. Python software for these calculations are under active development and are located at \cite{2Dgithub}.  
 
This expression is approximate because it does not correct for the protein footprint or any undulation-induced deviations of the membrane area.  The associated corrections are small compared to the membrane area and would shift the expected density for all species equally, without affecting the comparisons we perform here. For a given lipid species or class, analysis excluded any replicas in which fewer than 5 lipids of the species/class were in the leaflet at any point in the sampled simulation.

 \subsection{Calculation of the \newaffinity}
 
Although lipids to occupy clearly detectable hot-spots, binding to these sites are not straightforward to describe by a traditional two-state model. Lipids are chains that may partially occupy or fully occupy a site, and they may share a site with another lipid that is partly or fully occupying the site.  While the standard affinity can be determined from the probability of single occupancy, the \newaffinity{} is determined from the probability that a site is occupied by more beads than would be expected based on bulk density. 


For a given site, consider 
two probability distributions: the probability $P_{site}(n)$ of finding $n$ beads within the site and the probability $P_{bulk}(n)$ of finding $n$ beads within a region of equivalent area in the bulk, respectively.   For a lipid that has no affinity for this binding site, we expect $P_{site}(n) = P_{bulk}(n),$ while $P_{site}(n)$ should be right-shifted for a strong affinity and left-shifted in the presence of competition. 
We calculate the degree of right or left shift by first finding the number of beads $n_{peak}$ that corresponds to the peak of the density distribution in the bulk. As illustrated in Figure \ref{fig:PBT} C, we then integrate $P_{site}$ on both the left and right side of the threshold $n_{peak}$ to yield $P_{<}$ and $P_{>}$ respectively:
\begin{eqnarray}
    P_{<}& \equiv &\sum\nolimits_{n\le n_{peak}} P_{site}(n) \label{eq:Pl} \\
    P_{>}& \equiv &\sum\nolimits_{n>n_{peak}} P_{site}(n)  \label{eq:Pg}
\end{eqnarray}
Note that this step breaks the distribution into two macrostates on either side of the threshold, allowing clear analogy with a classic binary binding model.  
The free energy difference between the two macrostates is
\begin{equation}
\Delta G = -RT\ln\frac{P_{>}}{P_{<}}
\label{eq:dG}
\end{equation}
where R is the gas constant and T is temperature. We term this free energy difference the ``\newaffinity''.  In the special case of binary occupancy, 
\begin{eqnarray}
P_{site}(n)&= 
\begin{cases}
    (1+ K_{D}/[L])^{-1},& \text{if } n = 1\\
    (1+ [L]/K_{D})^{-1},& n=0
\end{cases}
\end{eqnarray}
where $K_{D}$ is the dissociation constant and $[L]$ is the ligand concentration. In a dilute solution the volume per ligand is typically much larger than the site volume, so $P_{bulk} (n)=1$ for $n=0$ and vanishes for all $n>0$, so $n_{peak} = 0$. Consequently, for this special case, $P_{<} = (1+ [L]/K_{D})^{-1}$ and $P_{>} = (1+ K_{D}/[L])^{-1}$. 
Then Equation \ref{eq:dG} reduces to the classic form for the chemical potential $RT\ln K_{D}-  RT\ln [L]$. 

\subsection{Binding Site Definition and Occupancy Calculations}

We consider two classes of site: intersubunit sites and M4 sites. Each \plgic~has ten of each site (five in the outer leaflet and five in the lower leaflet) for a total of twenty sites (Figure \ref{fig:PBT}B).  The boundaries for each site were drawn to correspond to the localized binding hot spots observed for heteroacidic membranes\cite{Woods2019}, and are non-overlapping. Inter-subunit sites include bins with angular components between the M1 and M3 alpha-helices of two adjacent subunits, and a radial component satisfying $10<r\leq32$\AA.  M4 sites include bins with complementary angular components (so that no sites overlap) falling within the M1 and M3 alpha-helices of a single subunit, and a radial component satisfying $10<r\leq44$\AA. For a full description of radial and angular dependencies, please see Table SI 4. 

In order to calculate $P_{site}(n)$, a distribution was taken across frames at 10 ns intervals. For any frame, the  beads of a given lipid or chain type were binned onto a fine polar grid with $\Delta r=$  4\AA and $\Delta \theta=$ $\frac{\pi}{25}$. The bins falling within the site boundaries were then summed to calculate the occupancy $n$.  This approach allowed for straightforward adjustment of site boundaries if needed without needing to re-bin the whole trajectory.  

\subsection{Calculation of Accessible Area}

Calculation of $P_{bulk}$ requires determining the accessible site area in order to calculate the densities in a bulk region of similar area. The area $A$ accessible to the lipids is the difference between the total site area $A_{tot}$ and the area $A_{ex}$ excluded by the protein: $A = A_{tot} - A_{ex}  $
$A_{tot}$ is straightforward to calculate by summing over the areas of the bins $i$ within the site boundaries: $A_{tot} = \sum_i r_i \Delta r_i  \Delta \theta_i$.  Calculating $A_{ex}$ is less straightforward, and although there are many possible ways to do this, for self-consistency we used the same tools from our primary analysis.  

In a single lipid membrane, $P_{site}(n)=P_{bulk}(n)$ as long as $P_{bulk}(n)$ is calculated using the proper area $A$.  We exploit this identity to calculate $A$ for each site, by running a single \nachr{} in pure di-palmitoyl phosphatidylcholine (DPPC) for $\sim 370$ns and determining the value of $A$ for each site such that $P_{site}(n)$ and $P_{bulk}(n)$ have the same peak.  These areas are reported in Table SI 4. 

 \begin{figure*}
	\center
	\includegraphics[width=\linewidth]{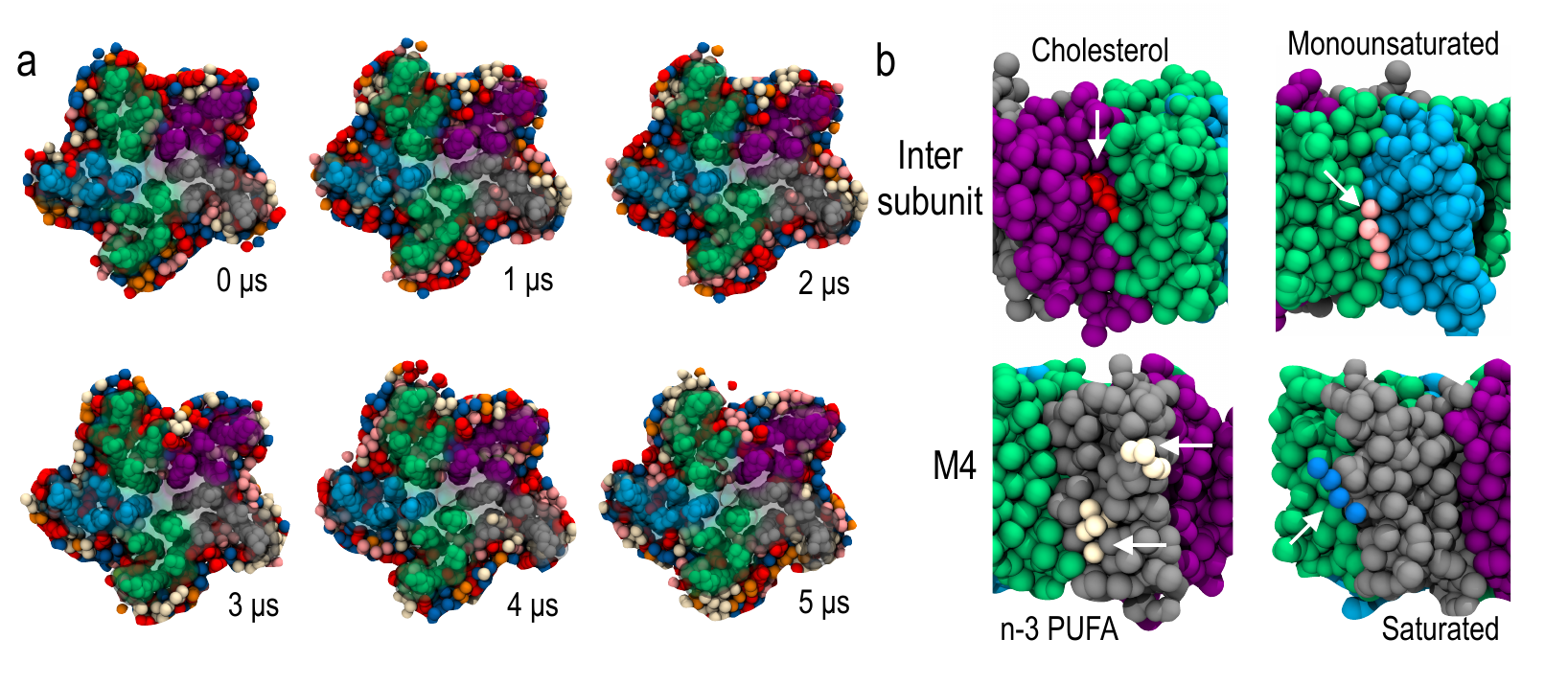}
	\caption{{A molecular perspective of coarse-grained simulation results}. a) Multiple frames from a single simulation replica over 5 $\mu s$.   The \nachr{} TMD is shown in surface representation and colored as in Figure 1. Cholesterol and acyl chains within 15 \AA~ of \nachr{} are shown as beads, and colored by chain type: saturated lipids: blue, monounsaturated lipids:orange, n-6 PUFAs:pink, n-3 PUFAs: beige, and cholesterol: red.  Each phospholipid color includes several lipid species of the same type, and simulations included a larger membrane and the ECD, which is not shown.  b) Representative poses of lipids for individual sites, colored as in A, but viewed from within the membrane looking at the TMD surface. Cholesterol selects for the intersubunit site while monounsaturated lipids have a particularly low affinity for this site. PUFAs select for the M4 site, while saturated lipids have a particularly low affinity.}
	\label{fig:trj}
\end{figure*}

\section{Results and Discussion}
\label{res}

\begin{figure*}[!h]
	\center
	\includegraphics[width=4in]{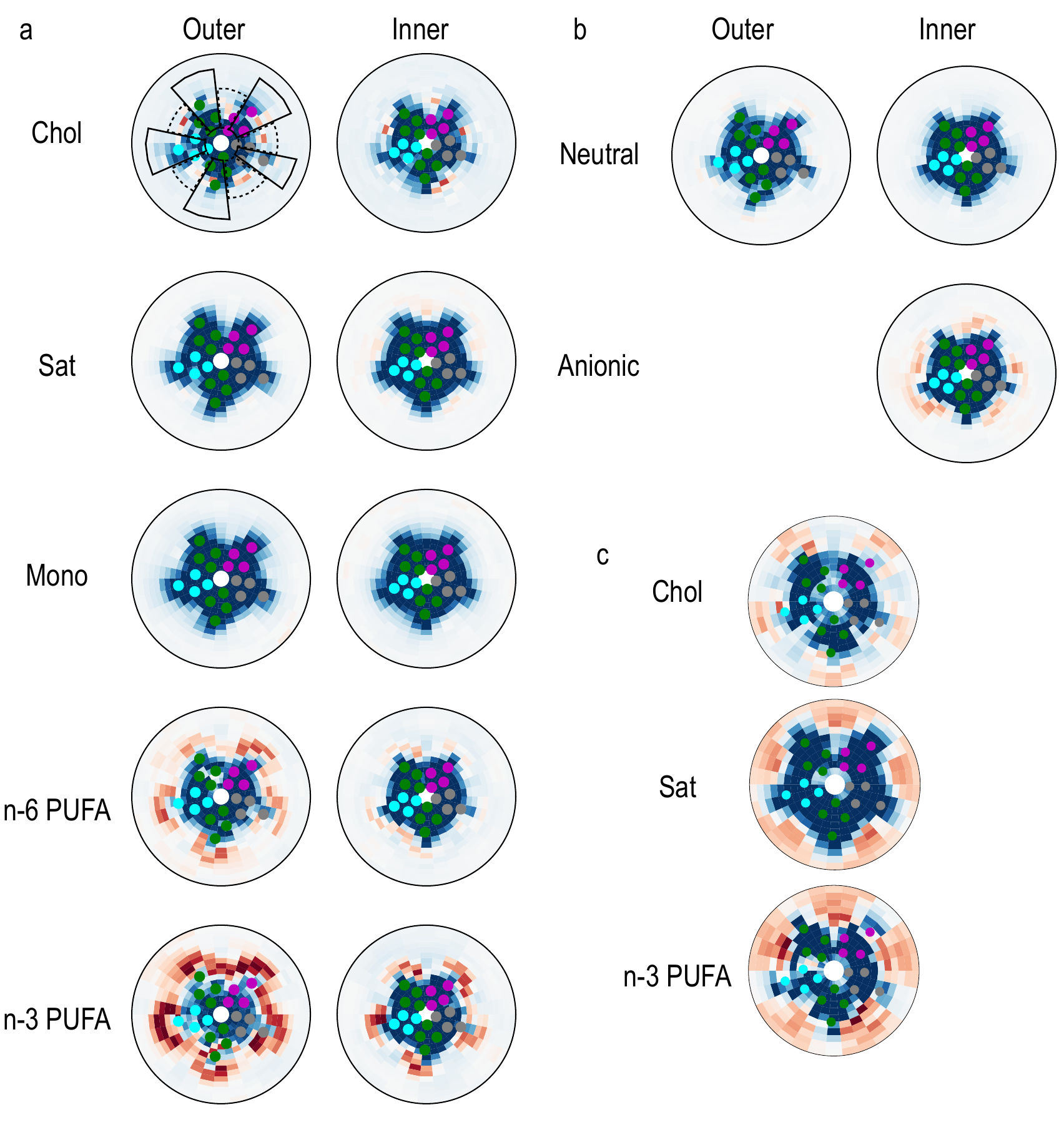}
	\caption{{Lipid density enrichment around a central singular \nachr.}  (a) and (b) Density enrichment $\tilde{\rho}_{a}$ for lipids in a neuronal membrane, calculated using eq \ref{eq:Rt} for both outer and inner leaflets, averaged over 10 replicas for 2.5 $\mu$s each. The maximum radius from the \nachr{} pore is 60 \AA. Depletion relative to a random mixture ($\log\tilde{\rho}_{a}< 0$) is blue while enrichment ($\log\tilde{\rho}_{a}> 0$) is red. Lipids are organized by acyl chain (a) or headgroup (b). Acyl chain density includes only the relevant chain of a heteroacidic lipid, while headgroup density includes the whole lipid.  Helices are represented as circles colored as in Figure 1. Intersubunit (solid line) and M4 (dashed line) site boundaries are marked.  (c) Equivalent analysis for \nachr{} in a model membrane of 2:2:1 n-3 PUFA:saturated:cholesterol, using previously published trajectories\cite{Woods2019}. } 
	\label{fig:acyl_map}
\end{figure*}

\subsection{Effect of acyl chain on site selectivity among neutral lipids}



\begin{figure*}[!h]
	\center
	\includegraphics[width=\linewidth]{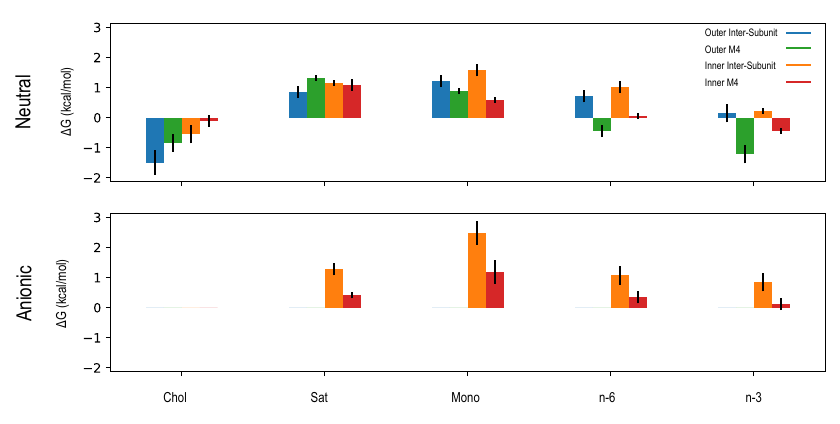}
	\caption{{\Newaffinities{} organized to reveal site selectivity.} The \newaffinities ($\Delta G$) are calculated using Equation \ref{eq:dG}, where error bars are the standard error (n=10 independent replicas).  \Newaffinities are colored by site; in the outer leaflet: intersubunit (blue) and  M4 (green), and for the inner leaflet: intersubunit (orange) and M4 (red). Values are separated by headgroup charge (rows) and acyl chain type (columns). More negative values indicate stronger affinities, while more positive values indicate more displacement of the lipid by other lipid species. Data incorporates 10 replicas averaging over the last half of the 5$us$ trajectory, with five-fold averaging over each type of pseudosymmetric site. Figure \ref{fig:lipidBar} has an alternate representation of the same data.  }  
	\label{fig:proBar}
\end{figure*}

Representative frames from a typical trajectory of boundary lipids are shown in Figure \ref{fig:trj}A, with representative poses shown in Figure \ref{fig:trj}B.  In order to quantitatively compare the lipid distributions for the native system to our previous model system, we plotted the enrichment of boundary density relative to bulk density on a two-dimensional polar heat map centered around the protein. This enrichment is shown in Figure \ref{fig:acyl_map}A for cholesterol and various acyl chains grouped by  saturation. Saturated and monounsaturated acyl chains are not significantly depleted or enriched in the boundary of the protein. Regions of cholesterol density are much more localized than in the model membrane (Figure \ref{fig:acyl_map}C) , with pockets of high enrichment very close to the protein and weak depletion in the remainder of the boundary region.  Both n-6 and n-3 PUFA chains yield five-fold symmetric enrichment around the M4 alpha-helices, as observed for n-3 PUFAs in the model membrane.  In the neuronal membrane, however, this enrichment is less well-defined and spreads into the intersubunit regions. In particular, additional pockets of significant enrichment are apparent in the $\beta-\delta$ subunit interface in the outer leaflet. The overall area of the regions of PUFA-enrichment decrease in the inner leaflet, where n-3 PUFAs are enriched around M4 helices, but n-6 PUFA density is not five-fold symmetric and has weak enrichment. Overall, the loss of definition in site boundaries diverges from the well-defined five fold enrichment for n-3 PUFAs we saw in model membranes\cite{Woods2019}. 

In order to reduce these distributions to affinities that are more straightforward to interpret, we calculated the \newaffinity{}  $\Delta G$ for various lipid species as defined in Eq. \ref{eq:dG}. We organize this information in two different ways: Figure \ref{fig:proBar} provides the ``lipid's perspective'' and is organized to identify the preferred site for a given lipid type (the lipids' ``site selectivity''), while Figure \ref{fig:lipidBar} provides the ``site's perspective'' and is organized to identify the most favorable lipids for a given site (the sites' ``lipid specificity''). 

We first consider site selectivity for neutral lipids. Affinities for neutral lipids and cholesterol are shown in Figure \ref{fig:proBar}A, where more negative values of $\Delta G$ indicate a stronger \newaffinity{} and more positive values indicate more displacement by other lipids.  Overall, as shown in Figure \ref{fig:proBar}A, saturated lipids have similar \newaffinities{} across all sites, which is consistent with the generally flat distribution observed in Figure \ref{fig:acyl_map}. Yet saturated lipids do yield a slightly stronger affinity for intersubunit sites, at least in the outer leaflet, which may drive the high amount of saturated enrichment observed at these sites in model membranes.  Outer leaflet monounsaturated lipids are slightly more unfavorable in intersubunit sites than M4 sites, and this difference grows in the inner leaflet. 

In contrast to saturated and monounsaturated lipids, PUFAs and cholesterol are highly selective for particular sites.  As shown in Figure  \ref{fig:proBar}A, neutral PUFAs have significantly stronger affinities for M4 sites than for innersubunit sites in the same leaflet.  Such selectivity is consistent with the PUFA enrichment density in Figure \ref{fig:acyl_map}A, where n-3 PUFAs can occupy most regions of the TMD but have particularly high levels of enrichment around M4. It is further consistent with our expectations from model membranes (Figure \ref{fig:acyl_map}C). Regardless of the site class, PUFAs favor the outer leaflet site over the inner leaflet site, but both sets of M4 sites are more favorable than both sets of intersubunit sites.  Conversely, cholesterol has significantly stronger affinities for innersubunit sites than for M4 sites, which is also consistent with the enrichment density in Figure \ref{fig:acyl_map}A and our expectations from model membranes (Figure \ref{fig:acyl_map}C). For cholesterol, however, the leaflet is a bigger determinant of affinity than the site; cholesterol has a stronger affinity for either outer leaflet site compared to either inner leaflet site.  

\subsection{Lipid preferences of intersubunit and M4 sites}
\begin{figure*}[!h]
	\center
	\includegraphics[width=\linewidth]{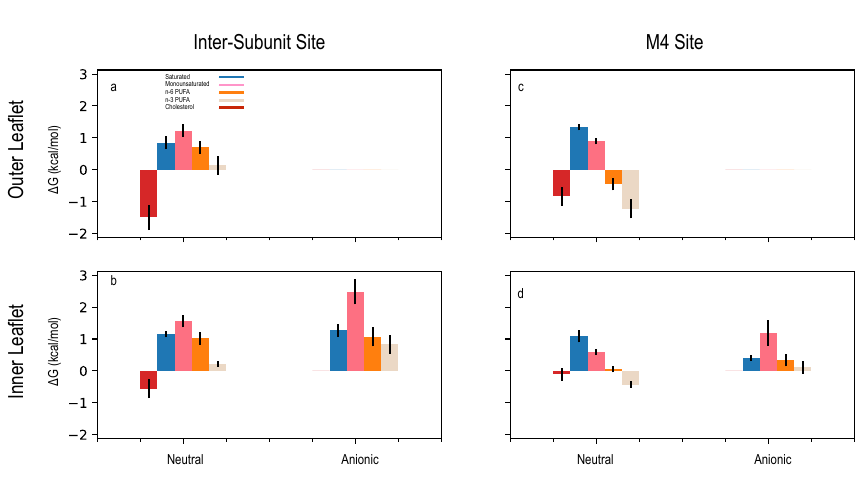}
	\caption{{\Newaffinities{} organized to reveal lipid preferences by site.} Data shown is identical but reorganized and recolored from Figure \ref{fig:proBar}. Here, \newaffinities are colored by chain type (Saturated:blue, Monounsaturated:pink, n-6 PUFAs:orange, n-3 PUFAs:tan, Cholesterol:red), and separated by leaflet (rows) and site (columns).} 
	\label{fig:lipidBar}
\end{figure*}

We now switch perspectives to considering which neutral lipids are most favorable for particular sites. As shown in Figure \ref{fig:lipidBar} A and B, intersubunit sites in both leaflets prefer cholesterol to phospholipids, which is expected based on the results from model membranes.  Upon visual inspection, this result may appear to diverge from the cholesterol polar density plots in neuronal membranes (Figure \ref{fig:acyl_map} A).  The present results show that while the overall footprint of cholesterol enrichment in (Figure \ref{fig:acyl_map} A) is small, this small region actually reflects a tight and persistently occupied binding site.  The highly right-shifted distributions for cholesterol are shown in Figure S1.  

PUFA chains yield affinities for the intersubunit site that are approximately $>$0.5 kcal/mol stronger than saturated lipid affinities (Figure \ref{fig:lipidBar} A and B), which was unexpected based on results from model membranes but is consistent with the corresponding enrichment density in Figure \ref{fig:acyl_map}A. More generally, neutral phospholipid affinities for intersubunit sites obey the following trend, from strongest to weakest: n-3 $>$ n-6 $>$ saturated $>$ monounsaturated. Thus, even though PUFA chains prefer M4 sites to intersubunit sites,  and saturated chains prefer intersubunit sites to M4 sites, PUFAs have a stronger affinity for either site type than do saturated lipids. 
 
For intersubunit sites, monounsaturated lipids have the weakest affinities ($>0.5$ kcal/mol), which may reflect a limited number of ways to pack the single kink of a monounsaturated chain in this concave site. In contrast, cholesterol and PUFAs are either small or highly flexible and may more easily pack across multiple sites. Saturated chains may pack parallel to the protein surface in these sites. 

As shown in Figure \ref{fig:lipidBar}C and D, M4 sites in both leaflets have the strongest affinity for n-3 PUFAs, and affinity weakens as acyl chain rigidity increases; from strongest to weakest the phospholipid affinities follow: n-3 $>$ n-6 $>$ monounsaturated $>$ saturated.  This is consistent with a role for PUFAs in minimizing unfavorable membrane deformations caused by the \plgic's conical-star shape.\cite{Brannigan2007,Kim1998,Dan1993,Goulian1996,Goulian1993,Fournier2015}  Surprisingly, cholesterol had a stronger affinity for M4 sites than any acyl chains other than n-3 PUFAs. Cholesterol is rigid, small, and has asymmetric sides (rough and smooth) which potentially allows it to embed between alpha-helices and compete with n-3 PUFAs for binding. Any cholesterol bound within the grooves of the subunit interface (as hypothesized based on atomistic simulations\cite{Brannigan2008} and observed in $\beta$ subunits  of \nachr{} (using coarse-grained simulations\cite{Sharp2019}), will also get counted within the M4 site.   


\subsection{Effect of Head Group Charge on Affinity Depends on Leaflet and Binding Site}
Figure \ref{fig:acyl_map}b compares the density enrichment for anionic headgroups with that of neutral headgroups. Data is shown for the inner leaflet only, because anionic lipids are not present in the outer leaflet at the start of simulations and few anionic lipids flip flop to the outer leaflet.

In the inner leaflet, the anionic lipids are expected to select for sites that are lined with basic amino acids, which are in different locations depending on subunit (Figure \ref{fig:aaa})  As shown in Figure \ref{fig:acyl_map}b, anionic lipids are generally enriched around the M3/M4 helices for the $\alpha_{\gamma}, \gamma,\delta,$ and $\beta$ subunits.  Anionic lipids are enriched at intersubunit sites and around M4 sites for all subunits but the $\alpha$ subunits. Non-$\alpha$ \nachr~ subunits have basic amino acids closer to M4 alpha-helices, as shown in Figure SI 2a.  We incorporate data from all five pseudo-symmetric sites to obtain the \newaffinities{} reported in Figure \ref{fig:proBar}B, which suggest that anionic lipids have significantly stronger affinities for M4 sites on average. The average anionic affinity difference between inter-subunit and M4 sites is $\sim -1.0$ kcal/mol, as shown in Figures \ref{fig:proBar}, \ref{fig:lipidBar}d and SI Table 2. Although the magnitude of the affinity difference varies with acyl chain saturation, the sign is unchanged.     
We now switch again to the ``site perspective'' to compare whether inner leaflet sites would prefer occupancy by anionic or neutral lipids. 
As shown in Figure \ref{fig:lipidBar}C, lipid affinity values for inter-subunit sites are either insensitive to charge (saturated or n-6 PUFA chains) or weaker for anionic lipids by at least 0.5 kcal/mol (monounsaturated and n-3 PUFA chains).  In comparison, at the M4 site, saturated chains in anionic lipids have significantly stronger affinities than those in neutral lipids (a difference of $\sim$0.5 kcal/mol). All other lipid chains attached to anionic headgroups have weaker affinities for the M4 site. The clear trend observed in neutral lipids (stronger affinities for more flexible acyl chains) is thus broken in anionic lipids because saturated anionic lipids are so favorable.  

In summary, we observe that binding sites have clear preferences for particular head group charge and acyl-chain saturation, suggesting \nachr~lipid occupancy to be driven in two steps, a ``coarse-sorting'' by head groups, and then ``fine-sorting'' by acyl-chains.  A neutral lipid will occupy \nachr's boundary region but acyl chains dictate where specific lipids occupy \nachr. Anionic lipids diverge from this pattern at the inner M4 site which has the strongest affinity for anionic lipids independent of saturation. 
\subsection{Role of Individual Lipid Headgroups in Determining Affinity}
Neutral and anionic are bulk terms that categorize numerous lipid head-groups by charge. 
To understand the role of the chemical distinctions between head groups of like charge, we broke the headgroup affinities down by headgroup species in Table \ref{tab:dGOuterHG}.  
In the outer leaflet, lipids contain a mixture of PE and PC headgroups. The small neutral PE head group has the strongest affinity across all headgroups for both inter-subunit and M4 sites, -0.2$\pm$0.3 and -1.1$\pm$0.2 kcal/mol respectively. The larger neutral PC headgroups are weaker than PE by $\sim > 0.5$ kcal/mol. In living cells, as in this neuronal membrane, PUFAs are more frequently tethered to PE than to PC or SM \cite{Isolated1969, Taguchi2010, Breckenridge1973,Ingolfsson2017b,Lorent2020}, so it is possible that this affinity simply reflects the high affinity of PUFA chains. However, even for identical chains, both experimental and simulation data \cite{Sharp2019} suggests stronger PE-ELIC than PC-ELIC interactions.  

Table \ref{tab:dGInnerHG} shows specific head group affinities in the inner leaflet. 
As in the outer leaflet, lipids with PE headgroups still have the strongest affinity of all lipids, but in the inner leaflet we are also able to distinguish affinities for anionic species. For the intersubunit site, PI, PS, and PC have similar affinities (within statistical error), and have significantly stronger affinities for these sites than the phosphoinositides (PIPS) PIP1, PIP2, PIP3, which have a significantly stronger affinity than phosphatidic acid  (PA). Thus, from strongest to weakest, PE$>$PI$\sim$PS$\sim$PC$>>$PIP1$\sim$PIP2$\sim$PIP3$>>$PA for the intersubunit site. In contrast, at the M4 site, more significant differences among moderate affinity headgroups emerge. PI has significantly stronger affinity than PS (a difference of 0.3$\pm$0.1 kcal/mol), and PS has a significantly stronger affinity than PC (a difference of 0.2$\pm$0.1 kcal/mol).
From strongest to weakest, PE$>$PI$>$PS$>$PC$>>$PIP1$\sim$PIP2$\sim$PIP3$\sim$PA for the M4 site.    



\begin{table}
	\caption{\Newaffinities () of neutral lipids for both sites in the outer leaflet, by head group.  Errors are standard errors (n=10 independent replicas). }
    \centering
    \begin{tabular}{|l||c|c|}
    \hline
	{} &   Intersubunit Sites&  M4 Sites\\
	{} & $\Delta$G (kcal/mol) & $\Delta$G (kcal/mol)\\
	\hline
	PE	&-0.2 $\pm$ 0.3& -1.1$\pm$0.2\\
	PC & 1.4$\pm$0.2&	1.1 $\pm$0.2\\
	\hline
    \end{tabular}
    \label{tab:dGOuterHG}
\end{table}

\begin{table}
	\caption{\Newaffinities ($\Delta$G) of neutral lipids for both sites in the inner leaflet, by head group. Values are sorted by strength of affinity for intersubunit sites. Errors are standard errors (n=10 independent replicas). }
    \centering
    \begin{tabular}{|l||c|c|}
    \hline
	{} &  Inner Inter Sites&  Inner M4 Sites\\
	{} & $\Delta$G (kcal/mol) & $\Delta$G (kcal/mol) \\
	\hline
	PE& 0.3$\pm$0.2& -0.1$\pm$0.1\\
	PI&0.9$\pm$0.3 &  0.2$\pm$0.1\\
	PS&1.0 $\pm$0.2	&  0.4$\pm$0.1\\
	PC &1.0$\pm$0.2 &	0.9$\pm0.1$ \\
	PIP3	&2.6$\pm$0.4	 &  1.8 $\pm$0.4\\
	PIP2	&2.8 $\pm$0.2	&  2.1$\pm$0.4\\
	PIP1	&2.4 $\pm$0.3	&  2.1$\pm$0.4\\
	PA	&3.0 $\pm$0.3	&  2.2$\pm$0.4\\
	\hline
    \end{tabular}
    \label{tab:dGInnerHG}
\end{table}

\section{Conclusions}

\label{con}

Using coarse-grained simulations of the \nachr{} within a quasi-neuronal membrane containing over thirty lipid species, we have observed spontaneous lipid binding and quantified lipid specificity for two types of sites in the protein TMD.  These two site classes represent the most concave (intersubunit site) and convex (M4 site) portions of the star-shaped \nachr~ and were initially observed as ``hot spots'' in our previous simulations\cite{Woods2019,Tong2019} of model membranes. Compared to classic ligand binding sites, these sites are superficial and have a large volume. The ``ligands'' occupying them are also non-traditional: lipids are flexible chain molecules that may only partially occupy the site and are likely to share the site with other partially-occupying ligands.  While our lab has developed promising alchemical approaches\cite{Salari2018} for calculating traditional affinities of atomistic lipids for more highly localized, well-defined sites, these hot spots required a different approach. Here we have proposed a softer ``\newaffinity'' for characterizing these affinities from spontaneous, unbiased coarse-grained simulations. While we restrict the use of this method here to \nachr, it should be straighforward to extend to any other transmembrane proteins with detectable regions of density enrichment. 

\begin{figure}
	\center
	\includegraphics[width=3.5in]{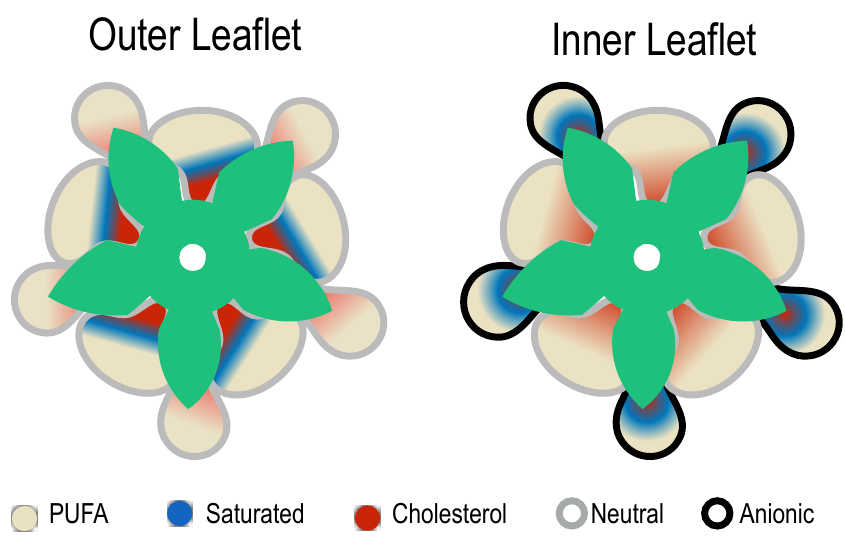}
	\caption{{Cartoon of expected boundary lipids for the \nachr~ in a native membrane for both leaflets}. Protein is shown in the center of both leaflets in a cyan floral shape. Grey and black outlines depict sites favorable for neutral and anionic lipids respectively. Fill color represents the lipids most likely to occupy each site (red: cholesterol, blue: saturated, beige: PUFA) and outline represents headgroup charge (gray: neutral, black: anionic).}
	\label{fig:sum}
\end{figure}

Our results are summarized graphically in Figure \ref{fig:sum}. Based on our results from model membranes, we had hypothesized that
 PUFAs would select for the convex M4 sites and that raft-forming lipids like cholesterol and saturated lipids would select for the concave inter-subunit sites. Overall, our results were consistent with this expectation.  Yet although lipids containing PUFAs do prefer the M4 site to the intersubunit site, their affinity for even the intersubunit sites are stronger than that of all other phospholipids. This result underscores the reliable partitioning of \nachr{} to PUFA-rich, liquid-disordered domains that we observed in homoacidic, domain forming membranes\cite{Sharp2019}, and suggests PUFAs may have been absent from the intersubunit site in heteroacidic membranes\cite{Woods2019} because of the constraints of the lipid topology. In the latter simulations, all lipids contained one saturated chain and one PUFA chain, so binding of the PUFA chain to its preferred M4 site requires the saturated chain to find the most favorable location nearby (in the intersubunit site) and may block binding of other PUFA chains to that site.  These constraints are relaxed in the native neuronal membrane, which has a more diverse lipid composition with multiple different chain pairings; about 6\% of the phospholipids in our simulated membranes contain no saturated chain at all. Nonetheless, our previous results\cite{Woods2019} using simplified binary heteroacidic/cholesterol membranes played a key role in identifying the natural site boundaries.  

As expected, within each leaflet cholesterol has the strongest affinity for the inter-subunit sites, although the affinity of cholesterol for the M4 sites was second only to that of n-3 PUFAs. Combined, these results are consistent with an overwhelming amount of evidence spanning four decades that suggests direct interactions between cholesterol and \nachr s, regardless of the phospholipid composition of the membrane.  One surprise for cholesterol was the role of the leaflet in determining affinity: cholesterol has a stronger affinity for either outer leaflet site compared to either inner leaflet site. This result may reflect competition with anionic saturated lipids in the inner leaflet, which would be consistent with multiple experiments\cite{Baenziger2000,Wenz2005,Hamouda2006,Thompson2020}, suggesting that anionic lipids can partially or fully compensate for a loss of cholesterol.  This result is also consistent with cholesterol embedded\cite{Brannigan2008} in the outer TMD (which has numerous gaps in the amino acid density) but not the inner TMD. 

Based on our results using ELIC\cite{Tong2019}, we had expected that anionic lipids would select for sites on the inner leaflet lined with basic residues.  In the homomeric ELIC, these residues are symmetricly-arranged, while in the heteromeric \nachr{} they vary by subunit(Figure \ref{fig:aaa}a), with the M4 site containing the most such residues on most subunits. The present results support that expectation: 
anionic lipids have a stronger affinity for M4 than inter-subunit sites.  

For both outer and inner leaflets, neutral lipids with smaller head groups (PE) have stronger affinity than the larger PC headgroup. It is unclear why PE is more favorable than other neutral lipids at this time, though this is consistent with previous work \cite{Sharp2019,Tong2019}, and the most straightforward explanation is that the smaller headgroup introduces fewer clashes with the protein TMD. 

Among anionic lipids in the inner leaflet, regardless of the site, PS and PI have an affinity greater than or equal to PC, and much greater than the other anionic lipid headgroups (PIP1,PIP2,PIP3, and PA).   The lipid headgroups PS and PI both have a charge of -1, while PA in the MARTINI forcefield\cite{DeJong2012} carries a charge of -2, and PIP1, PIP2, and PIP3 have charges of -3,-5, and -7. These results suggest that the inner leaflet sites select for monoanionic headgroups, while multianionic headgroups are highly unfavorable.  Due to the limitations of the coarse-grained model, future atomistic calculations are required to validate and understand the apparent preference of the M4 site for PI over PS.  

The present results highlight the utility of model membranes for developing hypotheses of specific lipid-protein interactions, and the need to test those hypothesis within more complex native membranes. The present results could be tested and aid in interpretation of experiments carried out in more complex membranes. For instance, we would expect that mutations of the basic residues facing the inner leaflet would reduce binding of saturated phospholipids with anionic headgroups, which would be replaced with bound cholesterol. We would also predict that if PUFAs cause gain of function via binding to the intersubunit site, this gain would be enhanced by replacing some heteroacidic lipids with homoacidic lipids while keeping the total fraction of PUFA chains constant. In general the present results provide valuable insight into how to predict lipid competition, which is one of the primarily challenges of interpreting experiments in complex membranes.  

\section*{Acknowledgments}
GB and LS were supported by the Busch Biomedical Foundation. This project was supported by generous allocation through the Rutgers University Office of Advanced Research Computing (OARC), which is supported by Rutgers University and the state of New Jersey. We are grateful to Dr. J{\'{e}}r{\^{o}}me H{\'{e}}nin for helpful input and suggestions.

\section*{Data Availability Statement}
The data that support the findings of this study are available from the corresponding author upon reasonable request. Scripts for polar density analysis and plotting scripts can be found on github: https://github.com/BranniganLab/densitymap. 






\bibliographystyle{model1-num-names}
\bibliography{Paper_Bibs}

\end{document}


\begin{titlepage}
   \begin{center}
      \huge\textbf{SUPPLEMENTARY MATERIAL}\\
      \LARGE{Spontaneous Lipid Binding to the Nicotinic Acetylcholine Receptor in a Native Membrane}\\

      Liam Sharp, Grace Brannigan
   \end{center}
   \vspace*{\stretch{2.0}}
   
\end{titlepage}

\renewcommand{\thefigure}{SI 1}
\begin{figure*}[!h]
	\center
	\includegraphics[width=\linewidth]{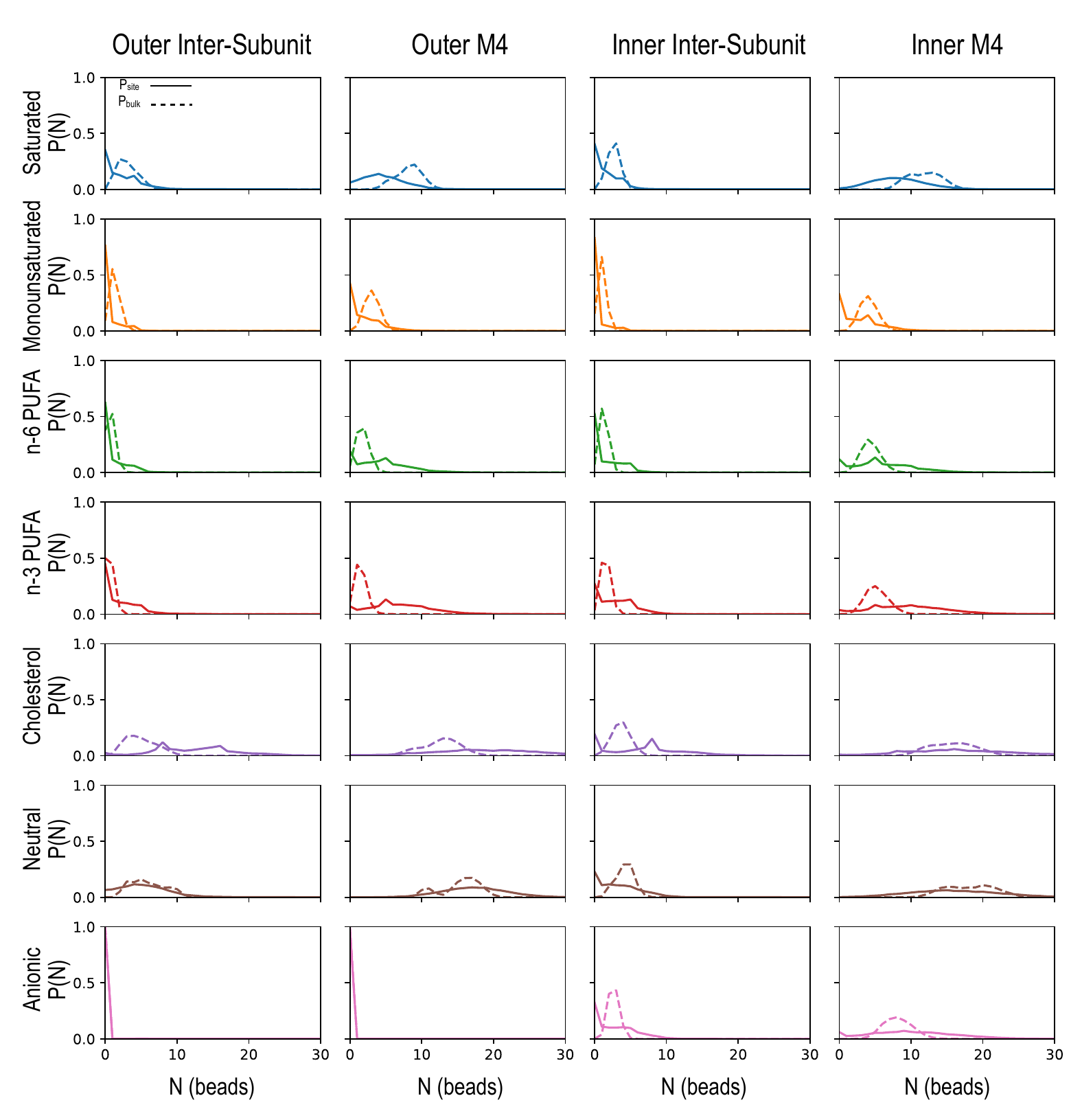}
	\caption{Probability distributions of acyl-chain saturations, including cholesterol, and head group charge. Solid lines represent $P_{site}$, the probability of a given number of beads found at occupancy site, averaged over both the course of the simulation and subunit sites. Dashed lines represent  $P_{bulk}$, the probability of a given number of beads in the bulk averaged over time. Bulk areas are square regions of equal area to occupancy sites.}
	\label{fig:lipidDist}
\end{figure*}

\renewcommand{\thetable}{SI 1}
\begin{table}
    \caption{Lipid ratios used for neuronal simulations grouped by head group.}
    \label{tab:rats}
    \centering

\begin{tabular}{|c||c|cc|}

\hline
Head Group & Lipids & Outer (\%) &{Inner (\%)} \\ \hline\hline
{}&CHOL                         & 44.3     & 40.67                         \\
\hline
PC &{} &30.5&15\\ \hline
{} &DPPC                         & 6.7      & 3.3                           \\
{} &DOPC                         & 2.8      & 1.4                           \\
{} &POPC                         & 11     & 5.4                         \\
{} &PFPC                         & 0.7      & 0.4                           \\
{} &PAPC                         & 5.9      & 2.9                           \\
{} &PUPC                         & 2.1      & 1.0                           \\
{} &OIPC                         & 0.7      & 0.4                           \\
{} &OUPC                         & 0.5      & 0.3                           \\
\hline
\hline
PE &{} &13.8&23.4\\ \hline
{} &POPE                         & 1.6      & 2.7                           \\
{} &PAPE                         & 4      & 6.7                           \\
{} &PUPE                         & 6.3      & 10.7                          \\
{} &OIPE                         & 0.2      & 0.3                           \\
{} &OAPE                         & 0.9      & 1.5                           \\
{} &OUPE                         & 0.9      & 1.5                          \\
\hline
\hline
SM &{} &11.3&2.5\\ \hline
{} &DPSM                         & 7.4      & 1.7                           \\
{} &PBSM                         & 1.4      & 0.3                           \\
{} &POSM                         & 0.9      & 0.2                           \\
{} &PNSM                         & 1.7     & 0.4                           \\
\hline
\hline
PS &{} &0.0&10.8\\ \hline
{} &DPPS                         & 0.0      & 0.5                           \\
{} &POPS                         & 0.0      & 2.7                           \\
{} &PAPS                         & 0.0      & 3.0                           \\
{} &PUPS                         & 0.0      & 3.8   			\\      
{} &OUPS                         & 0.0      & 0.8                           \\
\hline       
\hline          
PA &{} &0.0&0.4\\ \hline
{} &POPA                         & 0.0      & 0.1                          \\
{} &PAPA                         & 0.0      & 0.3                           \\
\hline
\hline
PI &{} &0.0&2.3\\ \hline
{} &POPI                         & 0.0      & 1.4                         \\
{} &PIPI                         & 0.0      & 0.6                           \\
{} &PAPI                         & 0.0      & 1.4                           \\
{} &PUPI                         & 0.0      & 2.3                           \\
\hline
\hline
PIPS &{} &0.0&1.5\\ \hline
{} &POP1                         & 0.0      & 0.2                           \\
{} &PAP1                         & 0.0      & 0.3                           \\
{} &POP2                         & 0.0      & 0.2                           \\
{} &PAP2                         & 0.0      & 0.3                           \\
{} &POP3                         & 0.0      & 0.2                           \\
{} &PAP3                         & 0.0      & 0.3                           \\
\hline
\end{tabular}
\end{table}

\renewcommand{\thefigure}{SI 2}

\begin{figure}
	\center
	\includegraphics[width=2in]{Anonic_AA.pdf}
	\caption{Charged amino acids in or near the TMD of two \plgic s and corresponding anionic enrichment. a) Structure of the TMD viewed from the extracellular side looking at the membrane, of \nachr~(top)\cite{Unwin2005} and ELIC\cite{Pan2012} (bottom). \nachr~ is colored as in Figure 1, ELIC is cyan. Basic amino acids are colored in blue while acidic amino acids are colored in red for both structures. b) Anionic polar density enrichment observed in ELIC, derived from \cite{Tong2019}. Grey circles represent center of mass of alpha-helices, red circles represent center of mass of alpha-helices with basic amino acids. }
	\label{fig:aaa}
\end{figure}
\renewcommand{\thetable}{SI 2}

\begin{table*}
    \caption{Affinities broken down by headgroup and acyl chain to reveal cross-correlation. }
    \centering
  \resizebox{\textwidth}{!}{  \begin{tabular}{|l||cc|cc|}
        
        \hline
        {} &  Outer Inter Sites&  Outer M4 Sites&  Inner Inter Sites&  Inner M4 Sites\\
        {} & $\Delta G$ (kcal/mol) & $\Delta G$ (kcal/mol) & $\Delta G$ (kcal/mol) & $\Delta G$ (kcal/mol) \\
        \hline
        
        CHOL &-1.5 $\pm$0.4& -0.8$\pm$0.3&    -0.6$\pm$0.3& -0.1$\pm$0.2\\
        Sat   &  0.7 $\pm$0.3	&  1.3 $\pm$0.2&     1.0 $\pm$	0.2&  1.1 $\pm$0.2 \\
        Mono   &    1.2 $\pm$0.2&  0.8$\pm$	0.2&     1.4 $\pm$0.1&  0.7 $\pm$0.2\\
        n-6   &         0.6 $\pm$0.2& -0.5 $\pm$0.1&0.3 $\pm$0.2& -0.3 $\pm$0.2\\
        n-3   &       -0.0 $\pm$0.4& -1.3 $\pm$0.3&    -0.2$0.2\pm$ & -0.9$\pm$0.2\\
        \hline
        Neutral &     0.4 $\pm$0.3& -0.4$\pm$0.2 &     0.5 $\pm$0.2&  0.5$\pm$0.2 \\
        Anionic &     2.4 $\pm$0.4&  2.4 $\pm$0.4&     0.3 $\pm$0.2& -0.3 $\pm$0.1\\
	\hline
        \hline
        Sat Neutral &  0.8 $\pm$0.2&  1.3 $\pm$0.1&     1.1$\pm$0.1 &  1.1 $\pm$0.2\\
        Mono Neutral & 1.2$\pm$0.2&  0.9 $\pm$0.1&     1.6 $\pm$0.2&  0.6 $\pm$0.1\\
        n-6 Neutral &  0.7 $\pm$0.2& -0.5 $\pm$0.2&     1.0 $\pm$0.2&  0.0 $\pm$0.1\\
        n-3 Neutral &  0.1 $\pm$0.3& -1.2$\pm$0.3& 0.2 $\pm$0.1& -0.4 $\pm$0.1\\
        \hline
        \hline
        Sat Anionic &&& 1.3$\pm$0.2 &  0.4$\pm$0.1\\
        Mono Anionic &&& 2.5 $\pm$0.4&  1.2 $\pm$0.4\\
        n-6 Anionic &&&1.1 $\pm$0.3&  0.3$\pm$0.2\\
        n-3 Anionic && &0.8 $\pm$0.3&  0.1 $\pm$0.2\\
        \hline

        \hline
        
    \end{tabular}}
    \label{tab:dGTab}
\end{table*}

\renewcommand{\thetable}{SI 3.1}
\begin{table}
    \caption{{Complete listing of affinities for intersubunit sites in the outer leaflet.}}
    \centering
    \tiny

\resizebox{\textwidth}{!}{ \begin{tabular}{| c || ccccc |}
\hline
    Lipids     & Outer $\alpha_{\gamma}-\beta$  & Outer $\beta-\delta$ & Outer $\delta-\alpha_{\delta}$  & Outer $\alpha_{\delta}-\gamma$  & Outer $\gamma-\alpha_{\gamma}$  \\
        \hline
        & ($\Delta G$ (kcal/mol)) & ($\Delta G$ (kcal/mol)) & ($\Delta G$ (kcal/mol)) & ($\Delta G$ (kcal/mol)) & ($\Delta G$ (kcal/mol)) \\
Lipid Species &&&&&\\
CHOL    &-0.9&-1.4&-1.5&-1.6&-2.0\\
DOPC    &3.5&3.4&2.2&2.8&2.3\\
DPPC    &3.3&2.1&2.3&1.4&2.0\\
DPPS    &4.1&4.1&4.2&3.9&4.2 \\
DPSM    &2.1&2.7&2.3&1.2&2.1\\
OAPE    &1.9&2.5&1.9&1.4&1.9\\
OIPC    &3.0&2.9&2.4&4.0&2.6\\
OIPE    &2.4&4.0&3.5&3.8&2.4\\
OUPC    &2.0&2.2&1.9&1.7&2.9\\
OUPE    &0.9&1.0&1.4&2.0&1.7\\
PAPC    &0.6&0.9&1.0&1.1&1.2\\
PAPE    &1.1&1.2&1.1&1.4&1.2\\
PBSM    &4.9&4.9&3.4&4.4&3.1\\
PFPC    &1.9&1.3&2.4&3.1&3.3\\
PNSM    &3.2&4.9&2.5&4.3&2.7\\
POPC    &2.2&1.8&1.7&2.0&1.6\\
POPE    &2.8&2.6&2.4&3.0&2.6\\
POSM    &3.3&2.6&3.8&2.9&3.4\\
PUPC    &1.4&1.5&1.9&1.0&1.9\\
PUPE    &0.0&0.3&0.1&-0.5&0.2\\
\hline
Head Groups 
PC      &0.9&0.4&1.0&0.7&1.1\\
PE      &-0.4&-0.3&-0.2&-0.3&0.4\\
SM      &1.9&2.5&2.2&1.2&2.0\\
\hline
Acyl-Chain Saturation 
Sat      &1.0&0.6&0.9&0.5&0.7\\
Monounsat      &1.1&1.1&1.1&1.3&1.4\\
n-6&0.4&0.5&0.8&0.7&0.7\\
n-3&-0.2&-0.2&0.0&-0.1&0.5\\
\hline
Head Group Charge&&&&&\\
Neutral &0.7&-0.1&0.7&0.2&0.5\\
\hline
Acyl-Chain Saturation by Charge 
Neutral Sat    &1.0&0.6&0.9&0.5&1.2\\
Neutral Monounsat    &1.1&1.2&1.1&1.3&1.4\\
Neutral n-6&0.4&0.5&0.8&0.7&1.1\\
Neutral n-3&-0.2&0.2&0.3&-0.1&0.5\\
\hline
\end{tabular}}
\end{table}

\renewcommand{\thetable}{SI 3.2}
\begin{table}
    \caption{Complete listing of affinities for M4 sites in the outer leaflet.}
    \centering
    \tiny

\resizebox{\textwidth}{!}{\begin{tabular}{| c || ccccc |}
\hline
     Lipids     & Outer $\alpha_{\gamma}$   & Outer $\beta$   & Outer $\delta$    & Outer $\alpha_{\delta}$   & Outer $\gamma$   \\
        \hline
        & ($\Delta G$ (kcal/mol)) & ($\Delta G$ (kcal/mol)) & ($\Delta G$ (kcal/mol)) & ($\Delta G$ (kcal/mol)) & ($\Delta G$ (kcal/mol)) \\
Lipid Species &&&&&\\
CHOL    &-1.0&-1.3&-0.8&-0.8&-0.2\\
DOPC    &1.8&2.1&1.5&1.5&1.9\\
DPPC    &1.3&2.0&1.7&2.2&1.8\\
DPSM    &1.6&1.5&1.7&1.7&2.0\\
OAPE    &1.5&1.3&1.3&0.9&1.1\\
OIPC    &2.2&1.8&2.8&2.1&2.2\\
OIPE    &2.0&2.2&3.1&2.6&2.8\\
OUPC    &1.7&1.5&1.5&1.5&1.3\\
OUPE    &0.6&0.8&0.8&0.9&0.9\\
PAPC    &0.2&0.3&0.5&0.3&0.2\\
PAPE    &0.4&0.3&0.1&0.4&0.4\\
PBSM    &2.7&3.0&3.5&2.7&2.9\\
PFPC    &2.0&1.3&1.9&1.8&1.7\\
PNSM    &2.3&2.3&2.5&2.5&2.6\\
POPC    &1.1&1.4&1.5&1.4&1.7\\
POPE    &1.9&1.9&1.7&1.7&2.3\\
POSM    &2.7&2.8&2.8&2.9&2.9\\
PUPC    &0.5&0.8&1.0&0.8&0.5\\
PUPE    &-0.3&-0.6&-0.6&-1.0&-0.9\\
\hline
Head Groups &&&&&\\
PC      &0.3&0.2&0.8&0.8&0.3\\
PE      &-0.7&-1.0&-1.3&-1.3&-1.2\\
SM      &1.5&1.6&1.8&1.8&1.9\\
\hline
Acyl-Chain Saturation &&&&&\\
Sat      &1.0&1.0&1.3&1.9&1.4\\
Monounsat      &0.8&0.7&0.7&0.7&1.1\\
n-6&-0.4&-0.6&-0.3&-0.5&-0.4\\
n-3&-1.1&-1.2&-1.0&-1.3&-1.8\\
\hline
Head Group Charge &&&&&\\
Neutral &-0.2&-0.9&0.0&0.0&-0.7\\
\hline
Acyl-Chain Saturation by Charge 
Neutral Sat    &1.0&1.0&1.3&2.0&1.4\\
Neutral Monounsat    &0.8&0.7&0.7&1.2&1.1\\
Neutral n-6&-0.4&-0.6&-0.3&-0.5&-0.4\\
Neutral n-3&-0.8&-1.2&-1.0&-1.3&-1.8\\
\hline
\end{tabular}}
\end{table}

\renewcommand{\thetable}{SI 3.3}
\begin{table}
    \caption{Complete listing of affinities for intersubunit sites in the inner leaflet.}
    \centering
    \tiny

\resizebox{\textwidth}{!}{\begin{tabular}{| c || ccccc |}
\hline
   Lipids     & Inner $\alpha_{\gamma}-\beta$  & Inner $\beta-\delta$  & Inner $\delta-\alpha_{\delta}$ & Inner $\alpha_{\delta}-\gamma$  & Inner $\gamma-\alpha_{\gamma}$  \\
        \hline
        & ($\Delta G$ (kcal/mol)) & ($\Delta G$ (kcal/mol)) & ($\Delta G$ (kcal/mol)) & ($\Delta G$ (kcal/mol)) & ($\Delta G$ (kcal/mol)) \\
Lipid Species &&&&&\\
CHOL    &-0.4&-0.4&-1.1&0.3&-1.2\\
DOPC    &2.4&3.1&2.9&2.6&2.9\\
DPPC    &1.9&2.7&2.1&2.3&2.1\\
DPPS    &4.4&4.2&3.5&4.9&5.3\\
DPSM    &2.5&1.7&2.1&2.6&2.5\\
OAPE    &2.3&2.5&1.9&1.9&3.1\\
OIPC    &3.1&2.9&3.8&3.0&3.4\\
OIPE    &4.9&3.3&3.5&3.8&3.3\\
OUPC    &3.1&3.2&2.3&2.7&2.1\\
OUPE    &2.4&1.4&1.5&1.8&1.7\\
OUPS    &2.4&2.8&1.5&2.2&2.1\\
PAP1&1.4&3.3&1.5&1.6&2.9\\
PAP2&1.9&3.9&            &1.1&2.1\\
PAP3&2.7&1.7&3.2&1.6&2.0\\
PAPA    &3.9&2.4&2.7&3.1&3.2\\
PAPC    &1.7&1.7&1.7&1.6&1.9\\
PAPE    &0.9&1.5&1.1&1.4&1.2\\
PAPI    &1.7&1.7&1.9&1.6&1.4\\
PAPS    &1.5&2.2&1.8&1.7&1.7\\
PBSM    &3.2&3.5&3.5&3.8&3.5\\
PFPC    &2.2&2.6&2.9&2.6&3.6\\
PIPI    &3.6&2.9&1.8&3.3&2.4\\
PNSM    &3.0&2.8&2.8&3.6&3.2\\
POP1&5.3&            &3.2&3.3&4.6\\
POP2&3.5&3.5&3.5&3.4&            \\
POP3&3.2&4.4&            &3.4&            \\
POPA    &            &            &5.3&5.3&       \\     
POPC    &2.4&1.7&2.0&2.0&1.8\\
POPE    &2.7&3.0&2.6&3.4&2.4\\
POPI    &2.4&2.9&2.5&2.4&2.3\\
POPS    &2.7&3.0&1.7&3.8&2.8\\
POSM    &3.9&4.1&2.7&4.3&3.4\\
PUPC    &2.0&1.7&1.7&2.1&2.0\\
PUPE    &0.1&0.2&0.4&0.5&0.4\\
PUPI    &1.0&1.0&1.4&1.1&0.7\\
PUPS    &1.2&1.7&1.0&1.5&1.6\\
\hline
Head Groups 
PC      &1.1&1.0&1.1&1.2&1.5\\
PE      &-0.1&0.3&0.3&0.4&0.6\\
SM      &2.4&1.7&1.9&2.6&2.5\\
PS      &0.8&1.4&0.5&1.1&1.4\\
PA      &3.7&2.5&2.6&3.1&3.3\\
PI      &0.9&0.9&1.0&0.9&1.0\\
PIP1&1.5&3.6&1.6&2.0&3.3\\
PIP2&2.1&4.0&4.1&1.3&2.4\\
PIP3&3.3&2.2&3.6&1.8&2.3\\
\hline
Acyl-Chain Saturation 
Sat      &0.7&1.1&0.9&1.1&1.3\\
Monounsat      &1.4&1.5&0.9&1.4&1.6\\
n-6&0.1&0.6&0.3&0.0&0.7\\
n-3&-0.3&0.0&-0.3&-0.2&0.1\\
\hline
Head Group Charge 
Neutral &0.4&0.7&0.5&0.5&0.5\\
Anionic &0.1&0.8&0.1&0.1&0.4\\
\hline
Acyl-Chain Saturation by Charge 
Neutral Sat    &1.0&1.1&1.1&1.3&1.3\\
Neutral Monounsat    &1.7&1.6&1.5&1.6&1.4\\
Neutral n-6&0.7&1.2&0.8&1.0&1.4\\
Neutral n-3&0.0&0.0&0.1&0.3&0.6\\
\hline
Anionic Sat     &1.0&1.6&1.1&1.3&1.3\\
Anionic Monounsat    &2.4&3.0&1.6&3.0&2.4\\
Anionic n-6&0.9&1.6&1.1&0.6&1.1\\
Anionic n-3&0.7&1.1&0.6&0.9&0.8\\
\hline
\end{tabular}}
\end{table}

\renewcommand{\thetable}{SI 3.4}
\begin{table}
    \caption{Complete listing of affinities for M4 sites in the inner leaflet.}
        \tiny
	\centering

\resizebox{\textwidth}{!}{\begin{tabular}{| c || ccccc |}
\hline	
     Lipids     & Inner $\alpha_{\gamma}$    & Inner$\beta$   & Inner $\delta$    & Inner $\alpha_{\delta}$    & Inner $\gamma$   \\
        \hline
        & ($\Delta G$ (kcal/mol)) & ($\Delta G$ (kcal/mol)) & ($\Delta G$ (kcal/mol)) & ($\Delta G$ (kcal/mol)) & ($\Delta G$ (kcal/mol)) \\ \hline 
Lipid Species &&&&&\\
CHOL    &-0.3&0.2&-0.4&-0.3&0.3        \\
DOPC    &1.6&1.4&1.8&1.5&1.6        \\
DPPC    &1.1&1.1&1.4&1.2&1.3        \\
DPPS    &2.2&2.6&2.9&2.6&3.1        \\
DPSM    &1.6&1.3&1.6&1.3&1.6        \\
OAPE    &1.4&1.3&1.5&1.2&1.4        \\
OIPC    &2.1&2.0&2.2&2.0&1.9        \\
OIPE    &2.2&2.8&2.7&2.1&2.4        \\
OUPC    &2.2&2.3&2.6&2.0&1.6        \\
OUPE    &1.2&1.0&1.0&0.9&0.8        \\
OUPS    &1.5&2.0&1.6&1.5&1.6        \\
PAP1&2.6&1.7&2.4&1.9&1.3        \\
PAP2&2.7&1.7&2.8&2.4&1.1        \\
PAP3&2.0&1.9&1.3&2.3&1.1        \\
PAPA    &2.6&1.9&2.4&2.6&2.0        \\
PAPC    &0.8&0.8&0.9&0.7&0.8        \\
PAPE    &0.4&0.4&0.5&0.3&0.4        \\
PAPI    &1.2&1.1&1.0&1.2&1.1        \\
PAPS    &0.8&0.9&1.1&1.0&0.7        \\
PBSM    &2.2&2.1&2.3&2.5&2.5        \\
PFPC    &2.2&2.0&2.1&1.9&2.1        \\
PIPI    &2.3&2.1&2.3&1.8&2.7        \\
PNSM    &2.4&2.1&2.1&1.9&2.5        \\
POP1&3.6&3.5&2.9&3.0&2.6        \\
POP2&3.9&2.6&2.4&3.0&2.6        \\
POP3&2.9&2.9&2.9&3.3&2.2        \\
POPA    &3.1&3.2&3.8&3.1&3.4        \\
POPC    &0.9&1.0&0.9&0.9&1.1        \\
POPE    &1.3&1.3&1.6&1.4&1.6        \\
POPI    &1.5&1.6&1.7&1.3&1.7        \\
POPS    &1.6&1.6&1.5&1.5&1.8        \\
POSM    &2.5&2.7&2.9&2.8&2.6        \\
PUPC    &1.5&1.2&1.5&1.1&1.0        \\
PUPE    &-0.1&-0.4&-0.1&-0.1&-0.3       \\
PUPI    &0.9&0.4&0.4&0.9&0.3        \\
PUPS    &0.9&0.7&0.5&0.8&0.7        \\
\hline
Head Groups &&&&&\\
PC      &0.6&0.7&0.8&0.7&0.6        \\
PE      &0.0&-0.2&-0.3&-0.1&-0.1       \\
SM      &1.4&1.4&1.4&1.1&1.6        \\
PS      &0.3&0.5&0.3&0.5&0.3        \\
PA      &2.5&1.8&2.4&2.5&2.0        \\
PI      &0.4&0.2&0.1&0.4&0.1        \\
PIP1&2.7&1.8&2.3&2.1&1.4        \\
PIP2&2.8&1.7&2.3&2.4&1.2        \\
PIP3&2.0&2.1&1.4&2.3&1.2        \\
\hline
Acyl-Chain Saturation &&&&&\\
Sat      &0.9&1.2&0.8&1.4&1.2        \\
Monounsat      &0.4&0.8&0.5&0.8&0.9        \\
n-6&-0.1&-0.4&-0.3&-0.1&-0.6       \\
n-3&-0.5&-1.0&-1.0&-0.5&-1.2       \\
\hline
Head Group Charge  &&&&&\\
Neutral &0.3&0.4&0.3&0.6&0.5        \\
Anionic &0.0&-0.3&-0.4&0.2&-0.7      \\
\hline
Acyl-Chain Saturation by Charge &&&&&\\
Neutral Sat    &0.9&1.2&0.8&1.2&1.4        \\
Neutral Monounsat    &0.5&0.7&0.6&0.4&0.8        \\
Neutral n-6&0.0&0.1&0.1&-0.1&0.1        \\
Neutral n-3&-0.2&-0.6&-0.4&-0.4&-0.6       \\
Anionic Sat     &0.3&0.5&0.1&0.7&0.3        \\
Anionic Monounsat    &1.1&1.2&1.2&1.2&1.2        \\
Anionic n-6&0.5&0.2&0.5&0.5&-0.1       \\
Anionic n-3&0.5&0.0&0.0&0.3&-0.1       \\\hline
\end{tabular}}
\end{table}

\renewcommand{\thetable}{SI 4.1}
\begin{table}
    \caption{Table SI Parameters used for calculating density threshold affinitiesfor intersubunit sites in the outer leaflet. Angular and radial boundaries are used to define sites for a given total area. Accessible areas are as described in Methods \textit{Binding Site Definition and Occupancy Calculations}. Methods \textit{Calculation of Accessible Area}}
    \centering
\resizebox{\textwidth}{!}{ \begin{tabular}{| c || ccccc |}
\hline
   {}     & Outer $\alpha_{\gamma}-\beta$  & Outer $\beta-\delta$ & Outer $\delta-\alpha_{\delta}$  & Outer $\alpha_{\delta}-\gamma$  & Outer $\gamma-\alpha_{\gamma}$  \\ \hline
  Angular Boundaries &$1.13\geq\theta\leq1.63$ rad&$0.37\geq\theta\leq6.16$ rad&$4.9\geq\theta\leq5.4$ rad&$3.64\geq\theta\leq4.15$ rad&$2.38\geq\theta\leq2.89$ rad\\
 Radial Boundaries &$10<r\leq32$\AA&$10<r\leq32$\AA&$10<r\leq32$\AA&$10<r\leq32$\AA&$10<r\leq32$\AA\\
  Total Area &301.59\AA$^2$&361.91\AA$^2$&301.59\AA$^2$&301.59\AA$^2$&301.59\AA$^2$\\
  Accessible Area &104.40\AA$^2$&50.89\AA$^2$&63.80&81.20\AA$^2$&34.80\AA$^2$\\

  \hline

\end{tabular} }
\end{table}

\renewcommand{\thetable}{SI 4.2}
\begin{table}
    \caption{Table SI Parameters used for calculating density threshold affinities for M4 sites in the outer leaflet. Angular and radial boundaries are used to define sites for a given total area. Accessible areas are as described in Methods \textit{Binding Site Definition and Occupancy Calculations}. Methods \textit{Calculation of Accessible Area}}
    \centering
\resizebox{\linewidth}{!}{ \begin{tabular}{| c || ccccc |}
\hline
 {}     & Outer $\alpha_{\gamma}$   & Outer $\beta$   & Outer $\delta$    & Outer $\alpha_{\delta}$   & Outer $\gamma$  \\ 
 \hline
   Angular Boundaries &$1.76\geq\theta\leq2.26$ rad&$0.5\geq\theta\leq1$ rad&$5.52\geq\theta\leq6.03$ rad&$4.27\geq\theta\leq4.78$ rad&$3.02\geq\theta\leq3.52$ rad\\
   Radial Boundaries &$10<r\leq44$\AA&$10<r\leq44$\AA&$10<r\leq44$\AA&$10<r\leq44$\AA&$10<r\leq44$\AA\\
  Total Area &703.72\AA$^2$&703.72\AA$^2$&703.72\AA$^2$&703.72\AA$^2$&703.72\AA$^2$\\
  Accessible Area &173.14\AA$^2$&117.29\AA$^2$&184.31\AA$^2$&195.48\AA$^2$&161.97\AA$^2$\\

  \hline

\end{tabular} }
\end{table}

\renewcommand{\thetable}{SI 4.3}
\begin{table}
    \caption{Table SI Parameters used for calculating density threshold affinities for intersubunit sites in the inner leaflet. Angular and radial boundaries are used to define sites for a given total area. Accessible areas are as described in Methods \textit{Binding Site Definition and Occupancy Calculations}. Methods \textit{Calculation of Accessible Area}}
    \centering
\resizebox{\linewidth}{!}{ \begin{tabular}{| c || ccccc |}
\hline
   {}     & Inner $\alpha_{\gamma}-\beta$  & Inner $\beta-\delta$  & Inner $\delta-\alpha_{\delta}$ & Inner $\alpha_{\delta}-\gamma$  & Inner $\gamma-\alpha_{\gamma}$  \\
   \hline
   Angular Boundaries &$1.38\geq\theta\leq1.88$ rad&$0.13\geq\theta\leq0.62$ rad&$5.15\geq\theta\leq5.65$ rad&$3.77\geq\theta\leq4.27$ rad&$2.64\geq\theta\leq3.01$ rad\\
   Radial Boundaries &$10<r\leq32$\AA&$10<r\leq32$\AA&$10<r\leq32$\AA&$10<r\leq32$\AA&$10<r\leq32$\AA\\
  Total Area &301.59\AA$^2$&241.27\AA$^2$&301.59\AA$^2$&241.27\AA$^2$&241.27\AA$^2$\\
  Accessible Area &68.29\AA$^2$&74.68\AA$^2$&68.29\AA$^2$&63.19\AA$^2$&40.21\AA$^2$\\

  \hline

\end{tabular} }
\end{table}

\renewcommand{\thetable}{SI 4.4}
\begin{table}
    \caption{Table SI Parameters used for calculating density threshold affinities for M4 in the inner leaflet. Angular and radial boundaries are used to define sites for a given total area. Accessible areas are as described in Methods \textit{Binding Site Definition and Occupancy Calculations}. Methods \textit{Calculation of Accessible Area}}
    \centering
\resizebox{\linewidth}{!}{ \begin{tabular}{| c || ccccc |}
\hline
   {}     & Inner $\alpha_{\gamma}$    & Inner$\beta$   & Inner $\delta$    & Inner $\alpha_{\delta}$    & Inner $\gamma$\\   \hline
   Angular Boundaries &$2.01\geq\theta\leq2.51$ rad&$.75\geq\theta\leq1.26$ rad&$5.78\geq\theta\leq6.16$ rad&$4.4\geq\theta\leq5.03$ rad&$3.14\geq\theta\leq3.64$ rad\\
   Radial Boundaries &$10<r\leq44$\AA&$10<r\leq44$\AA&$10<r\leq44$\AA&$10<r\leq44$\AA&$10<r\leq44$\AA\\
  Total Area &703.72\AA$^2$&844.46\AA$^2$&703.72\AA$^2$&844.46\AA$^2$&844.46\AA$^2$\\
  Accessible Area &211.69\AA$^2$&273.88\AA$^2$&205.97\AA$^2$&302.41\AA$^2$&273.88\AA$^2$\\
  \hline

\end{tabular} }
\end{table}
\clearpage
\printbibliography